\documentclass[11pt,a4paper]{article}
\pdfoutput=1
\usepackage{jcappub}

\usepackage[normalem]{ulem}
\usepackage{comment}
\usepackage{mathtools}
\usepackage{braket}
\usepackage{bm}
\usepackage{subcaption}

\definecolor{dgreen}{rgb}{0.05, 0.60, 0.05}

\begin{document}


\title{Supermassive black hole formation from Affleck-Dine mechanism with suppressed clustering on large scales}
\author[a]{Kentaro Kasai}
\author[a,b]{, Masahiro Kawasaki}
\author[c]{, Kai Murai}
\author[a]{, \\Shunsuke Neda}
\affiliation[a]{ICRR, University of Tokyo, Kashiwa, 277-8582, Japan}
\affiliation[b]{Kavli IPMU (WPI), UTIAS, University of Tokyo, Kashiwa, 277-8583, Japan}
\affiliation[c]{Department of Physics, Tohoku University, Sendai, 980-8578, Japan}

\abstract{
We study a primordial black hole (PBH) formation model based on the framework of the inhomogeneous Affleck-Dine (AD) mechanism, which can explain the seeds of supermassive black holes (SMBHs).
This model, however, predicts strong clustering of SMBHs that is inconsistent with the observation of angular correlation of quasars.
In this paper, we propose a modified model that can significantly reduce the PBH clustering on large scales by considering a time-dependent Hubble-induced mass during inflation. 
The quasar angular correlation is suppressed by the large Hubble-induced mass in the early stage of inflation while the small Hubble-induced mass in the late stage leads to the AD field fluctuations large enough for PBH formation as in the original model. 
As a result, the modified scenario can successfully explain the seeds of SMBHs.
}


\emailAdd{kkasai@icrr.u-tokyo.ac.jp}
\emailAdd{kawasaki@icrr.u-tokyo.ac.jp}
\emailAdd{kai.murai.e2@tohoku.ac.jp}
\emailAdd{neda@icrr.u-tokyo.ac.jp}

\begin{flushright}
    TU-1229
\end{flushright}

\maketitle

\section{Introduction}
\label{sec: intro}

Supermassive black holes (SMBHs) are considered to reside in the centers of almost all massive galaxies~\cite{Kormendy:1995er,Magorrian:1997hw,Richstone:1998ky}.
Furtheremore, the pulsar timing array experiments, NANOGrav~\cite{NANOGrav:2023hfp}, EPTA~\cite{EPTA:2023fyk} with InPTA data, PPTA~\cite{Reardon:2023gzh}, and CPTA~\cite{Xu:2023wog}, recently reported the detection of stochastic gravitational waves that may be produced by inspiraling binaries of SMBHs.
The astrophysical origin of such SMBHs has been considered to be the accretion and merger of stellar black holes.
However, the observations by the James Webb Space Telescope~\cite{Gardner:2006ky} suggest the existence of SMBHs with masses of $\sim 10^6\,\text{--}\,10^{10}M_\odot$ at high redshifts up to $z \sim 10$~\cite{Onoue:2022goe,Kocevski:2023,Barro:2023,Ubler:2023,CEERSTeam:2023qgy,Harikane:2023,Maiolino:2023zdu,Bogdan:2023ilu,Maiolino:2023bpi,Goulding:2023gqa}.
The super-Eddington accretion is required for stellar black holes to grow into SMBHs at such high redshifts.
Thus, it is challenging to explain such SMBHs by astrophysical processes~\cite{Woods:2018lty}.

Another possibility for the origin of SMBHs is primordial black holes (PBHs), i.e., black holes formed from large density fluctuations in the early universe~\cite{Zeldovich:1967lct,Hawking:1971ei,Carr:1974nx}.
In contrast to stellar black holes, PBHs can be formed with an initial mass as large as $M \gtrsim 10^4 M_\odot$ and can be the seeds of SMBHs.
However, the formation of such heavy PBHs requires large density fluctuations on the corresponding scales $k\sim 10^{4\,\text{--}\,5}\,\mathrm{Mpc}^{-1}$ and induces the $\mu$-distortion of the CMB spectrum.
In particular, if the density fluctuations follow the Gaussian distribution, PBHs with $M \gtrsim 10^4 M_\odot$ are severely constrained from the non-observation of the $\mu$-distortion~\cite{Chluba:2012we,Kohri:2014lza}.
On the other hand, if PBHs are formed from highly non-Gaussian fluctuations, this constraint can be weakened or evaded~\cite{Nakama:2017xvq,Unal:2020mts,Gow:2022jfb} (see also Refs.~\cite{Sharma:2024img,Byrnes:2024vjt} for the effects of non-Gaussianity on the $\mu$-distortion and its implications for the constraint on PBHs).
The PBH formation models that produce such highly non-Gaussian density fluctuations are proposed in Refs.~\cite{Dolgov:1992pu,Dolgov:2008wu,Hasegawa:2017jtk,Hasegawa:2018yuy,Kawasaki:2019iis,Kawasaki:2021zir,Kasai:2022vhq,Garriga:1992nm,Nakama:2016kfq,Garcia-Bellido:2017aan,Deng:2017uwc,Kitajima:2020kig,Kasai:2023ofh,Hooper:2023nnl,Kasai:2023qic}.

In this paper, we consider the PBH formation from the inhomogeneous Affleck-Dine (AD) mechanism~\cite{Dolgov:1992pu,Dolgov:2008wu,Hasegawa:2017jtk,Hasegawa:2018yuy,Kawasaki:2019iis,Kawasaki:2021zir,Kasai:2022vhq} (see also Ref.~\cite{Dolgov:2023ijt}). 
In this scenario, the AD field fluctuates spatially by the effect of quantum diffusion during inflation.
Due to the evolution of the potential for the AD field after inflation, the universe is separated into large and small-field regions.
The small-field regions occupy most of the universe and have negligible field values.
On the other hand, in the large-field regions,  the AD fields are large enough for the AD mechanism to produce large baryon or lepton asymmetry. 
The large-field regions are rare because they are produced from a large field tail of the distribution.
In such regions, the oscillation of the AD field experiences spatial instabilities and fragments into non-topological solitons called Q-balls~\cite{Coleman:1985ki,Kusenko:1997zq,Kusenko:1997si,Enqvist:1997si,Kasuya:1999wu,Kasuya:2000wx,Enqvist:2000gq}.
Since Q-balls behave as non-relativistic matter, the density contrast between inside and outside the large-field regions grows during the radiation-dominated era, and the large-field regions finally collapse into PBHs at the horizon reentry. 
Since the large-field regions occupy a tiny volume fraction of the universe, their density fluctuations give a negligible contribution to the $\mu$-distortion, and hence avoid the stringent constraint from the CMB.
This type of model, however, inevitably induces the clustering of PBHs due to the large-scale perturbations of the AD field.
Such strong clustering of PBHs is constrained by observations of isocurvature perturbations~\cite{Kawasaki:2021zir,Kasai:2023ofh} and angular correlations of quasars~\cite{Shinohara:2021psq,Shinohara:2023wjd}.
These constraints can be relaxed by suppressing the large-scale fluctuations of the field related to the PBH formation~\cite{Kasai:2023qic}.

Therefore, in this paper, we propose an extension of the AD scenario for the formation of the seeds of SMBHs.
In this scenario, we consider a change of the Hubble mass during inflation, which is realized, for example, in multi-inflaton models.
In particular, we suppose that a Hubble mass is large in the first stage of inflation and small in the subsequent second stage, which can suppress large-scale perturbations of the AD field while keeping the advantages of the original model.
Consequently, we find that the constraint from clustering is significantly relaxed and that the origin of the SMBHs can be explained in this scenario.

The rest of this paper is organized as follows.
In Sec.~\ref{sec: review}, we briefly review the PBH formation through the inhomogeneous AD mechanism.
In Sec.~\ref{sec: modif}, we introduce the modified setup.
Then, we discuss the observational constraints on PBHs and their clustering in Sec.~\ref{sec: PBH obs} and those on the L-ball scenario in Sec.~\ref{sec: L-ball}.
Sec.~\ref{sec: summary} is devoted to the summary and discussion of our results.

\section{Review of PBH formation from inhomogeneous AD mechanism}
\label{sec: review}

First, we briefly review PBH formation from the inhomogeneous AD mechanism following Ref.~\cite{Kasai:2022vhq}.

\subsection{Inhomogeneous Affleck-Dine mechanism}
\label{subsec: AD mechanism}

This PBH formation scenario is based on the AD mechanism~\cite{Affleck:1984fy,Dine:1995kz}, in which a baryon or lepton charge is generated by the dynamics of the AD field $\phi$ during and after inflation. 
The AD field is one of the flat directions in the scalar potential in the minimal supersymmetric standard model (MSSM) and has a baryon or lepton charge.
The potential of the AD field arises from the supersymmetry breaking and Planck-suppressed effects.
In particular, we consider the potential for the AD field given by 
\begin{align}
    V(\phi)
    =
    \left\{
    \begin{aligned}
        &(m_\phi^2 + c H^2)|\phi|^2
        + V_\mathrm{NR} + V_\mathrm{A}
        &\quad \text{(During~inflation)}
        \\
        &(m_\phi^2 - \tilde{c} H^2)|\phi|^2
        + V_\mathrm{NR} + V_\mathrm{A} + V_\mathrm{T}
        &\quad \text{(After~inflation)}
    \end{aligned}
    \right.
    \ ,
\end{align}
where $m_\phi$ is a soft SUSY breaking mass, $H$ is the Hubble parameter, and $c$ and $\tilde{c}$ are positive dimensionless constants.
In addition to the mass term, we have the non-renormalizable term, $V_\mathrm{NR}$, A-term, $V_\mathrm{A}$, and thermal potential, $V_\mathrm{T}$, given by
\begin{align}
    V_\mathrm{NR}
    &=
    |\kappa|^2 \frac{|\phi|^{2(n-1)}}{M_\mathrm{Pl}^{2(n-3)}}
    \label{eq : V NR}
    \ ,
    \\
    V_\mathrm{A}
    &=
    \kappa a_M \frac{m_{3/2}\phi^n}{n M_\mathrm{Pl}^{n-3}}
    + \mathrm{h.c.}
    \ ,
    \\
    \label{eq:thermal_pot}
    V_\mathrm{T}
    &=
    \left\{
    \begin{aligned}
        &d_1 T^2 |\phi|^2
        & \quad (|\phi| \lesssim T)
        \\
        &d_2 T^4 \log \left( \frac{|\phi|^2}{T^2} \right)
        & \quad (|\phi| \gtrsim T)
    \end{aligned}
    \right.
    \ ,
\end{align}
respectively.
Here, $\kappa$, $a_M$, and $d_{1,2}$ are dimensionless constants, $n \, (\geq 4)$ is an integer determined by the choice of the flat direction, $M_\mathrm{Pl} \simeq 2.4 \times 10^{18}$\,GeV is the reduced Planck mass, $m_{3/2}$ is the gravitino mass, and $T$ is the cosmic temperature after inflation.
Notice that the A-term, $V_\mathrm{A}$, breaks the $U(1)$ symmetry associated with baryon or lepton number conservation. 

In the conventional AD mechanism, we assume $c < 0$, and $\phi$ has a nonzero field value during inflation.
After inflation, the soft SUSY breaking mass overcomes the negative Hubble mass term, and the AD field starts to oscillate.
At the same time, $\phi$ is kicked in the phase direction by $V_\mathrm{A}$, and the baryon or lepton charge is generated.

In the PBH formation scenario, we consider $c>0$, and then AD field $\phi$ tends to settle down to the origin.
However, since $\phi$ acquires quantum fluctuations, $\phi$ has a non-zero field value whose distribution is determined by the Hubble parameter during inflation and e-folds.
In this scenario, $\tilde{c}>0$ is also assumed, and the Hubble-induced mass term becomes negative just after inflation.
At the same time, the thermal bath is produced by the inflaton decay, which induces the thermal potential. 
We further assume that the effective mass due to the thermal potential is larger than the Hubble-induced mass around the origin ($d_1 T^2 > \tilde{c}H^2$), and the AD field is stabilized there.
Since $V_\mathrm{T}$ has a logarithmic shape for $\varphi \equiv |\phi| \gtrsim T$, it becomes subdominant for a large field value.
Then, another local minimum of $V(\phi)$ arises at
\begin{align}
    \varphi
    \sim
    \varphi_m
    \equiv
    \left(
    \frac{\sqrt{\tilde{c}}HM_\mathrm{Pl}^{n-3}}{\lvert\kappa\rvert}
    \right)^\frac{1}{n-2}
    \ ,
\end{align}
due to the balance between the negative Hubble mass term and $V_\mathrm{NR}$.
Consequently, $V(\phi)$ comes to have multiple local minima just after inflation.

\begin{figure}[t]
    \centering
    \includegraphics[width=.8\textwidth ]{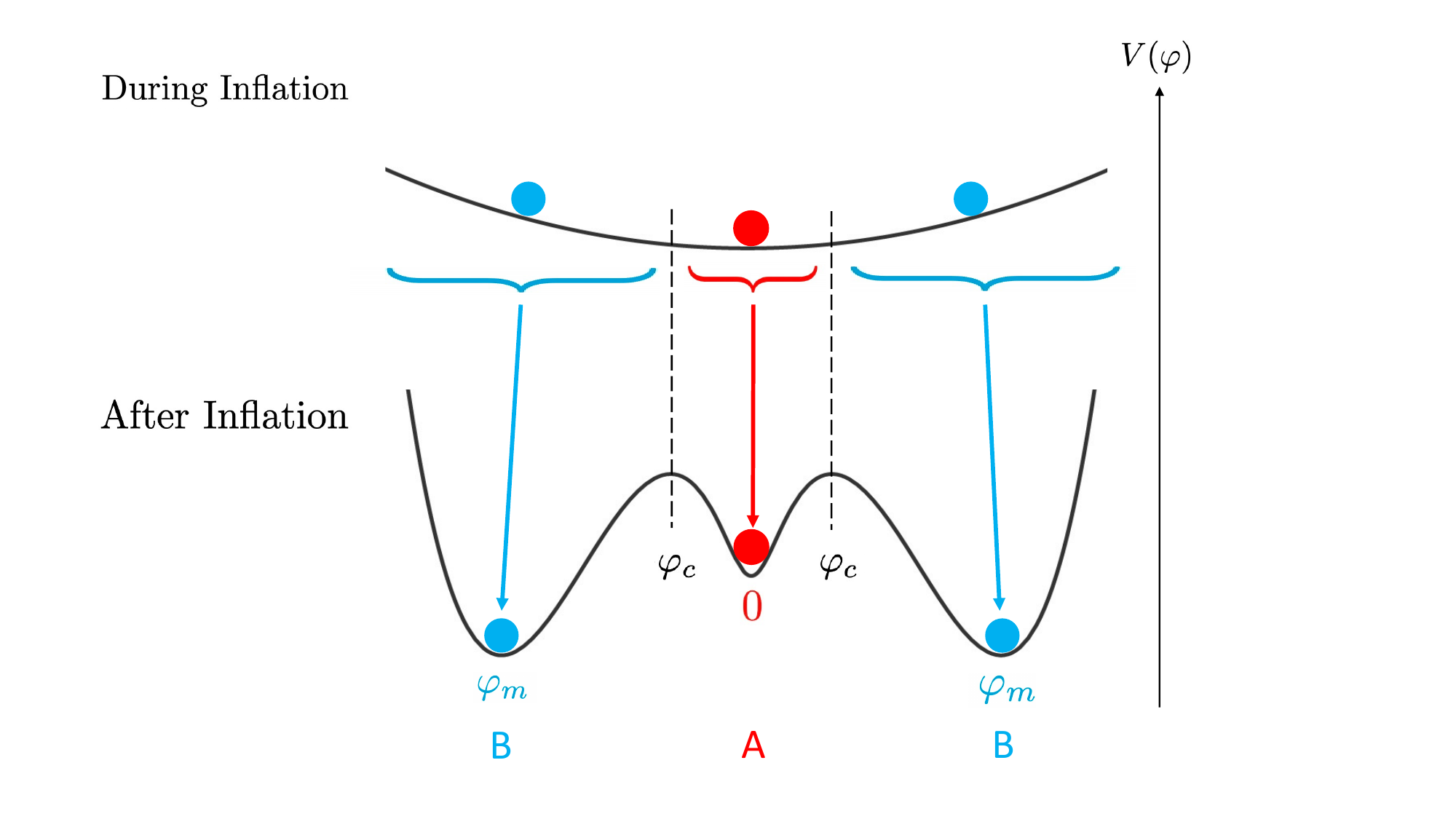}
    \caption{%
        Schematic view of the dynamics of the AD field during and after inflation.
        \textit{Upper side}: During inflation, the AD field coarse-grained over a Hubble patch diffuses in the complex plane and takes different values in different Hubble patches.
        \textit{Lower side}: Just after inflation, the multi-vacuum structure is realized. 
        In the regions with $\varphi \lesssim \varphi_c$, $\phi$ rolls down to vacuum A, and almost no lepton asymmetry is generated.
        On the other hand, if $\varphi \gtrsim \varphi_c$, $\phi$ classically rolls down to vacuum B, and those regions are identified as high lepton bubbles later.
    }
    \label{fig: potential}
\end{figure}

The local maximum between those two minima is calculated as
\begin{align}
    \varphi
    \sim
    \varphi_c
    \equiv
    \sqrt{\frac{d_2}{\tilde{c}}}
    \frac{T^2}{H}
    \ ,
\end{align}
from the Hubble-induced mass and the thermal potential. Here, we used $\log(\varphi^2/T^2)=\mathcal{O}(1)$. 
Note that this model requires $\varphi_c\gtrsim H_{\rm{end}}$, where $H_{\rm{end}}$ denotes the Hubble parameter at the end of inflation such that the origin, $\varphi = 0$, is stabilized after inflation.
If $\varphi$ is larger than $\varphi_c$ at the end of inflation, $\phi$ settles down to $\varphi = \varphi_m$, which we call ``vacuum B''.
In vacuum B, the AD mechanism works as in the conventional case, and the baryon or lepton charge is generated.
On the other hand, if $\varphi < \varphi_c$, $\phi$ settles down to $\varphi = 0$, which we call ``vacuum A'', and the AD mechanism does not work.
We show the schematic view of the dynamics of the AD field in Fig.~\ref{fig: potential}.
Hereafter we consider that a lepton charge is generated in the AD mechanism.\footnote{
If a baryon charge is generated in the AD mechanism, the inhomogeneous baryon density spoils the success of the big bang nucleosynthesis as shown in Ref.~\cite{Kasai:2022vhq}.}

After inflation, the negative Hubble-induced term decreases, and vacuum B is destabilized.
Then, the AD field starts to oscillate, and the lepton asymmetry is generated.
Assuming that $\varphi_c$ is larger than the typical amplitude of the fluctuation of $\phi$ during inflation, the regions with $\varphi>\varphi_c$ at the end of inflation become rare.
Thus, the resultant lepton asymmetric regions are also rare, and we call them high lepton bubbles.
The lepton-to-entropy ratio generated in the bubbles is given by 
\begin{align}
    \eta_l
    \equiv
    \frac{n_l}{s}
    \simeq 
    \epsilon \frac{T_\mathrm{RH}m_{3/2}}{H_\mathrm{osc}^2}
    \left( \frac{\varphi_\mathrm{osc}}{M_\mathrm{Pl}} \right)^2
    \ ,
    \label{eq: bubble asymmetry}
\end{align}
where $n_l$ is the lepton number density, $s$ is the total entropy density, $\epsilon$ is an efficiency parameter of the asymmetry generation, $T_\mathrm{RH}$ is the reheating temperature, and the subscripts ``osc'' denote quantities when the AD field starts to oscillate.
Hereafter we take $\epsilon = 0.1$.

\subsection{PBH formation}
\label{subsec: PBH formation}

Inside the high lepton bubbles, the AD field oscillates with a large amplitude.
Depending on the potential shape, the AD field experiences spatial instabilities and forms L-balls~\cite{Kawasaki:2002hq}, i.e., Q-balls with a lepton charge.
Since L-balls behave as non-relativistic matter, the density contrast between inside and outside the bubbles grows as the universe expands during the radiation-dominated era.
The energy per unit entropy inside the bubble is written as
\begin{align}
    Y_Q^\mathrm{in}
    \equiv 
    \frac{\rho_Q^\mathrm{in}}{s}
    =
    \omega_Q \eta_l
    \ ,
    \label{eq: YQin}
\end{align}
where $\rho_Q^\mathrm{in}$ is the energy density of the L-balls inside the bubble, and $\omega_Q$ is the energy of an L-ball per unit lepton charge.
Then, the density contrast is obtained as 
\begin{align}
    \label{eq:density_contrast}
    \delta 
    \equiv 
    \frac{\rho^\mathrm{in} - \rho^\mathrm{out}}{\rho^\mathrm{out}}
    \simeq 
    \frac{\rho_Q^\mathrm{in}}{\rho_\mathrm{rad}}
    =
    \frac{4 Y_Q^\mathrm{in}}{3 T}
    \ ,
\end{align}
where $\rho^\mathrm{in}$ and $\rho^\mathrm{out}$ are the total energy density inside and outside the bubbles, respectively, and $\rho_\mathrm{rad}$ is the energy density of the background universe.
More details on the properties of L-balls are discussed in Sec.~\ref{sec: L-ball}.

If the density contrast is large enough at the horizon reentry of the bubbles, PBHs are formed.
The threshold value of the density contrast evaluated on the comoving slice is~\cite{Harada:2013epa}
\begin{align}
    \delta_c 
    \simeq 
    \frac{3(1 + w)}{5+ 3w} \sin^2 \left( \frac{\pi\sqrt{w}}{1 + 3w} \right)
    \ ,
\end{align}
where $w$ is the parameter of the equation of state inside the bubbles given by 
\begin{align}
    w 
    =
    \frac{p^\mathrm{in}}{\rho^\mathrm{in}}
    \simeq 
    \frac{p^\mathrm{out}}{\rho^\mathrm{in}}
    =
    \frac{1}{3(1 + \delta)}
    \ .
\end{align}
Combining $\delta > \delta_c$ and the above formulae, we obtain the condition for PBH formation as 
\begin{align}
    \delta \gtrsim 0.40
    \ .
\end{align}
From Eq,~\eqref{eq:density_contrast}, we obtain the maximum temperature for PBH formation as 
\begin{align}
    T_c
    \simeq 
    3.4 Y_Q^\mathrm{in}
    \ .
    \label{eq:temp_max}
\end{align}

For later convenience, we relate the temperature at the horizon reentry to the corresponding wavenumber $k$.
From the Friedmann equation,
\begin{align}
    \frac{k}{a}
    \simeq 
    H
    =
    \sqrt{\frac{\pi^2 g_*}{90}} \frac{T^2}{M_\mathrm{Pl}}
    \ ,
\end{align}
and 
the entropy conservation,
\begin{align}
    g_{*s} T^3 a^3 = g_{*s0} T_0^3
    \ ,
\end{align}
we obtain the relation between $k$ and $T$ as 
\begin{align}
    T
    &=
    \sqrt{\frac{90}{\pi^2 g_*}} 
    \left( \frac{g_{*s}}{g_{*s0}} \right)^{1/3}
    \frac{M_\mathrm{Pl} k}{T_0}
    \nonumber \\
    &\simeq 
    3.9\,\mathrm{MeV}
    \left( \frac{g_*}{10.75} \right)^{-1/2}
    \left( \frac{g_{*s}}{10.75} \right)^{1/3}
    \left( \frac{k}{4.5\times 10^4\,\mathrm{Mpc}^{-1}} \right)
    \ ,
    \label{eq:PBH_T_k}
\end{align}
where $a$ is the scale factor, $g_*$ and $g_{*s}$ are the relativistic degrees of freedom for the energy and entropy densities, respectively, and the subscript ``0'' denotes quantities at the present time.
Here, we used $g_{*s0} = 43/11$, $T_0 = 2.725$\,K, and $a_0 =1$.

The PBH mass, defined as $M$, is given by the background horizon mass at PBH formation~\cite{Kopp:2010sh,Carr:2014pga}.
Thus, we can also relate $k$ to $M$ as
\begin{align}
    M
    &\simeq 
    \frac{4\pi}{3} \frac{\rho_\mathrm{rad}}{H^3}
    =
    4\pi M_\mathrm{Pl} \sqrt{\frac{\pi^2 g_*}{90}}
    \left( \frac{g_{*s0}}{g_{*s}} \right)^{2/3}
    \left( \frac{T_0}{k} \right)^{2}
    \nonumber \\
    &\simeq 
    1.0\times 10^4 M_\odot
    \left( \frac{g_*}{10.75} \right)^{1/2}
    \left( \frac{g_{*s}}{10.75} \right)^{-2/3}
    \left( \frac{k}{4.5\times 10^4\,\mathrm{Mpc}^{-1}} \right)^{-2}
    \ .
    \label{eq:PBH_mass_k}
\end{align}
From Eqs.~\eqref{eq:temp_max}, \eqref{eq:PBH_T_k}, and \eqref{eq:PBH_mass_k}, we obtain the minimum PBH mass as
\begin{align}
    M 
    \gtrsim
    M_c
    \simeq 
    1.3\times 10^4 M_\odot 
    \left( \frac{g_*}{10.75} \right)^{-1/2}
    \left( \frac{Y_Q^\mathrm{in}}{1\,\mathrm{MeV}} \right)^{-2}
    \ .
    \label{eq: PBH formation mass}
\end{align}
The bubbles that reenter the horizon at $T < T_c$ become PBHs, while the other bubbles remain as they are.
To discuss the seed PBHs of SMBHs, we focus on $M_c=10^4 M_\odot$ below.
From Eqs.~\eqref{eq: bubble asymmetry}, \eqref{eq: YQin}, and \eqref{eq: PBH formation mass}, we determine $\varphi_\mathrm{osc}$ for given $M_c, T_\mathrm{RH},$ and $m_{3/2}$.

\subsection{PBH mass spectrum}
\label{subsec: PBH mass spectrum}

Next, we consider the size distribution of the high lepton bubbles and the PBH mass spectrum.
The distribution of the bubbles is determined by quantum fluctuations of the AD field during inflation.
The fluctuations of the AD field coarse-grained over the Hubble scale are described by the probability distribution, $P(N, \phi)$.
Here, $N$ is the e-folding number during inflation defined by 
\begin{align}
    N(k) \equiv \ln\left( \frac{k}{k_0} \right)
    \ ,
\end{align}
where $k$ is the wavenumber that exits the horizon at $N$, and $k_0 \simeq 2.24 \times 10^{-4}$\,Mpc$^{-1}$ is the scale of the current observable universe.
In other words, we define $N$ so that the current horizon scale exits the horizon at $N = 0$ during inflation. 
The probability distribution follows the Fokker-Planck equation
\begin{align}
    \frac{\partial P(N,\phi)}{\partial N}
    =
    \sum_{n=1,2}\frac{\partial}{\partial \phi_n}
    \left[\frac{\partial V(\phi)}{\partial \phi_n}\frac{P(N,\phi)}{3H_I^2}
    +\frac{H_I^2}{8\pi^2}\frac{\partial P(N,\phi)}{\partial \phi_n}\right],
    \label{eq: Fokker-Planck 1}
\end{align} 
where $H_I$ is the Hubble parameter during inflation, which we assume to be constant, and $\phi_n\,(n=1,2)$ is the canonical real scalar field defined by
\begin{align}
    \phi
    =
    \frac{\phi_1+i\phi_2}{\sqrt{2}}
    .
    \label{eq: phi}
\end{align} 
We approximate the potential of the AD field by a Hubble-induced mass term,
\begin{align}
    V(\phi)=cH_I^2{\lvert\phi\rvert}^2.
    \label{eq: potential}
\end{align} 
To simplify the presentation, we use dimensionless fields defined by
\begin{align}
    \tilde{\phi} 
    \equiv 
    \tilde{\phi}_1 + i \tilde{\phi}_2
    \ , \quad 
    \tilde{\phi}_1 
    \equiv 
    \frac{\sqrt{2} \pi \phi_1}{H_I}
    \ , \quad 
    \tilde{\phi}_2 
    \equiv 
    \frac{\sqrt{2} \pi \phi_2}{H_I}
    \ .
\end{align}
Then, the Fokker-Planck equation becomes
\begin{align}
    \frac{\partial \tilde{P}(N,\tilde{\phi})}{\partial N}
    =
    c' \tilde{P}(N,\tilde{\phi}) 
    +
    \sum_{n=1,2} \left[
        \frac{c' \tilde{\phi}_n}{2}
        \frac{\partial \tilde{P}(N,\tilde{\phi})}{\partial \tilde{\phi}_n}
        +\frac{\partial^2 \tilde{P}(N,\tilde{\phi})}{\partial \tilde{\phi}_n^2}
    \right],
    \label{eq: Fokker-Planck dimless}
\end{align} 
where $c' \equiv 2 c/3$.
Here, $\tilde{P}$ is defined by
\begin{align}
    \tilde{P}(N, \tilde{\phi})
    \equiv 
    \left( \frac{H_I}{\sqrt{2}\pi} \right)^2
    P(N, \phi)
    \ ,
\end{align}
which satisfies the normalization of 
\begin{align}
    \iint \mathrm{d} \tilde{\phi}_1 \mathrm{d} \tilde{\phi}_2 \,
    \tilde{P}(N, \tilde{\phi})
    = 
    1
    \ .
\end{align}
Setting the initial condition as $\tilde{P}(N_i, \tilde{\phi}) = \delta^2(\tilde{\phi}-\tilde{\phi}_i)$, we obtain the solution as~\cite{Kawasaki:2021zir}
\begin{align}
    \tilde{P}(N,\tilde{\phi} ; c', N_i, \tilde{\phi}_i)
    =
    \frac{1}{2\pi\sigma^2(N-N_i, c')}
    \exp \left[
        -\frac{1}{2\sigma^2(N-N_i, c')}
        \Big\lvert 
        \tilde{\phi} - e^{-\frac{c'}{2}(N-N_i)} \tilde{\phi}_i
        \Big\rvert^2
    \right]
    \ ,
    \label{eq: itit}
\end{align}
where
\begin{align}
    \sigma^2(N,c')
    \equiv 
    \frac{1-e^{-c'N}}{2c'}
    \ .
\end{align}
In the following, we assume $N_i = 0$ and $\tilde{\phi}_i = 0$ for simplicity.

If $\tilde{\varphi} \equiv |\tilde{\phi}|$ is larger than the threshold value $\tilde{\varphi}_c = 2\pi \varphi_c/H_I$ at the end of inflation, bubbles will be formed.
Regarding that the super-horizon modes of the AD field obey the classical motion, we define an effective threshold value during inflation as
\begin{align}
    \tilde{\varphi}_{c,\mathrm{eff}}(N)
    \equiv 
    e^{\lambda
    (N_\mathrm{end} - N)}
    \tilde{\varphi}_c
    \ ,
\end{align}
where $N_\mathrm{end}$ is the e-folding number at the end of inflation, and
\begin{align}
    \lambda
    \equiv 
    \frac{3}{2}\left(1 - \sqrt{ 1 - \frac{2}{3}c' } \right)
    \ .
\end{align}
Note that $\lambda \simeq c'/2$ for $c' \ll 1$. 
Thus, we can regard that a region finally becomes a high lepton bubble after inflation if $\tilde{\varphi}$ coarse-grained over that region is larger than $\tilde{\varphi}_{c,\mathrm{eff}}(N)$ at the horizon exit.
From this observation, we can estimate the volume fraction of the bubbles on larger scales than $k^{-1}$ as 
\begin{align}
    B(N_k)
    =
    \int_{\tilde{\varphi}_{c,\mathrm{eff}}(N_k)}^{\infty}
    \mathrm{d} \tilde{\varphi} \,
    \bar{P}(N_k, \tilde{\varphi}; c', 0, 0)
    \ ,
    \label{eq: vol frac}
\end{align}
where $N_k \equiv N(k)$, and $\bar{P}$ is the probability distribution for $\tilde{\varphi}$ given by 
\begin{align}
    \bar{P}(N, \tilde{\varphi} ; c', N_i, \tilde{\varphi}_i)
    \equiv 
    \int_0^{2\pi} \mathrm{d} \theta \,
    \tilde{\varphi}
    \tilde{P}(N,\tilde{\varphi} e^{i \theta} ; c', N_i, \tilde{\varphi}_i)
    \ .
\end{align}
By differentiating the volume fraction with respect to $N_k$, we obtain the size distribution of the bubbles by 
\begin{align}
    \beta_N(N_k)
    \equiv 
    \frac{\partial B(N_k)}{\partial N_k}
    \ .
\end{align}
From $M \propto k^{-2} \propto e^{-2N}$, we can translate $\beta_N$ into the PBH formation rate per $\ln M$ as
\begin{align}
    \beta(M)
    \equiv 
    \frac{1}{2}\beta(N) \Theta(M - M_c)
    \ .
\end{align}
Here, $\Theta$ is the Heaviside step function.
Finally, we obtain the present energy density ratio of PBHs to the total dark matter as
\begin{align}
    f_\mathrm{PBH}
    \equiv 
    \frac{\Omega_\mathrm{PBH}}{\Omega_c}
    =
    \int \mathrm{d}(\ln M)\,
    \beta(M) \frac{T(M)}{T_\mathrm{eq}} 
    \frac{\Omega_m}{\Omega_c}
    \ ,
\end{align}
where $\Omega_\mathrm{PBH}$, $\Omega_c$, and $\Omega_m$ are the current density parameters of the PBHs, dark matter, and non-relativistic matter, respectively, $T(M)$ is the formation temperature of PBHs with a mass of $M$, and $T_\mathrm{eq}$ is the cosmic temperature at matter-radiation equality.

\subsection{PBH correlation}
\label{subsec: PBH correlation}

We can also estimate the two-point correlation of PBHs from the probability distribution.
In the following, we consider PBHs with a certain mass $M$ assuming a monochromatic mass distribution for simplicity.
We then define the PBH density fluctuation by
\begin{align}
    \delta_\mathrm{PBH}(\bm{x})
    \equiv 
    \frac{\rho_\mathrm{PBH}(\bm{x})}{\bar{\rho}_\mathrm{PBH}}
    - 1
    =
    \frac{1}{\bar{n}_\mathrm{PBH}}
    \sum_i \delta^{(3)}(\bm{x} - \bm{x}_i)
    - 1
    \ ,
\end{align}
where $\rho_\mathrm{PBH}$ and $n_\mathrm{PBH}$ are the energy and number densities of PBHs, the barred density denotes spatially averaged one, and $\bm{x}_i$ denotes the position of the $i$-th PBH.
From the above equation, the two-point correlation function is written as
\begin{align}
    \langle \delta_\mathrm{PBH}(0) \delta_\mathrm{PBH}(\bm{x})\rangle 
    &=
    \left\langle 
        \sum_{i,j} \frac{\delta^{(3)}(-\bm{x}_i) \delta^{(3)}(\bm{x} - \bm{x}_j)}{\bar{n}_\mathrm{PBH}^2}
        -
        \frac{\sum_i \delta^{(3)}(-\bm{x}_i) + \sum_j \delta^{(3)}(\bm{x} - \bm{x}_j)}{\bar{n}_\mathrm{PBH}}
        + 1
    \right\rangle
    \nonumber \\
    &=
    \sum_{i} \frac{\delta^{(3)}(\bm{x})}{\bar{n}_\mathrm{PBH}}
    +
    \sum_{i \neq j} \frac{\delta^{(3)}(-\bm{x}_i) \delta^{(3)}(\bm{x} - \bm{x}_j)}{\bar{n}_\mathrm{PBH}^2}
    - 1
    \nonumber \\
    &\equiv
    \sum_{i} \frac{\delta^{(3)}(\bm{x})}{\bar{n}_\mathrm{PBH}}
    +
    \xi(\bm{x})
    \ ,
\end{align}
where the bracket denotes the average over the positions of PBHs.
While the first term describes the Poisson fluctuation, $\xi$ is the reduced PBH correlation function describing the PBH clustering, which vanishes if PBHs are not correlated.

To evaluate the correlation function, we consider the probability distribution of the bubble formation at two points $x$ and $y$ separated by a comoving distance $L(=2\pi/k)$ given by 
\begin{align}
    \beta_{N2}(N_L, N_x, N_y)
    =
    &\frac{\partial^2}{\partial N_x \partial N_y}
    \int_0^{\infty} \mathrm{d} \tilde{\varphi}_L
    \bar{P}(N_k, \tilde{\varphi}; c', 0, 0)
    \nonumber \\
    &\times \int_{\tilde{\varphi}_{c,\mathrm{eff}}(N_x)}^{\infty}
    \mathrm{d} \tilde{\varphi}_x \,
    \bar{P}(N_x, \tilde{\varphi}_x; c', N_L, \tilde{\varphi}_L)
    \int_{\tilde{\varphi}_{c,\mathrm{eff}}(N_y)}^{\infty}
    \mathrm{d} \tilde{\varphi}_y \,
    \bar{P}(N_y, \tilde{\varphi}_y; c', N_L, \tilde{\varphi}_L)
    \ ,
\end{align}
where $N_x$ and $N_y$ are the e-folding numbers corresponding to the PBHs formed at $x$ and $y$, respectively, and $N_L$ is the e-folding number when $L$ exits the horizon.
Using $\beta_{N2}$, we obtain the correlation function as
\begin{align}
    \xi(L)
    =
    \frac{\beta_{N2}(N_L, N(M), N(M))}{\beta_{N}(N(M))^2}
    - 1
    \ .
\end{align}
In this scenario, the PBHs are strongly clustered even on large scales, which leads to large isocurvature perturbations and angular correlations.
Thus, the scenario is stringently constrained from the observations of CMB and angular correlations of quasars~\cite{Shinohara:2021psq,Shinohara:2023wjd} depending on the mass and abundance.
In particular, the latter excludes the scenario explaining seed PBHs for the SMBH as shown in Sec.~\ref{sec: PBH obs}.

\section{Modified model with double inflation}
\label{sec: modif}

In this section, we introduce a modification of the scenario described in the previous section and discuss the suppression of the correlation function.

\subsection{Setup}
\label{sec: setup}

The PBH correlation comes from the fluctuation of the AD field on scales larger than the PBH scale.
Thus, we expect that, if the quantum diffusion of the AD field is suppressed in the earlier stage of inflation compared to the later stage, the correlation function is also suppressed.
We can realize such a situation by considering the time dependence of the Hubble mass.
As an example, we consider double inflation~\cite{Silk:1986vc}.
In double inflation, two scalar fields work as the inflaton successively.
Thus, the coefficient of the Hubble mass is also naturally different in the first and second stages of inflation.
In the following, we assume that $c'$ changes during inflation as
\begin{align}
    c'=
    \begin{cases}
    c'_1 & N < N_\ast \\
    c'_2 & N \geq N_\ast
    \end{cases}
    \ ,
\end{align}
where $N_\ast$ denotes the e-folding number at the transition of inflationary stages.
For simplicity, we also assume that $H_I$ is constant during inflation although this is not the case in the realistic situation.
The possible effects of the time evolution of the Hubble parameter will be discussed later.

\subsection{Probability density for the modified setup}
\label{sec: modified setup}

\begin{figure}[t]
    \centering
    \includegraphics[width=.75\textwidth ]{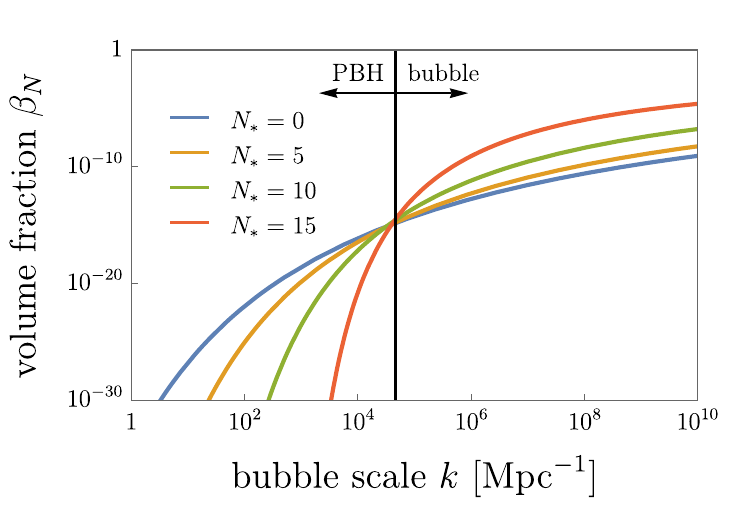}
    \caption{%
        Volume fraction of lepton bubbles per $\ln k$
        for different values of $N_\ast$. 
        The horizontal axis denotes the bubble size in terms of the wave number.
        The lepton bubbles on the left side of the black line form PBHs, and those on the right side remain as bubbles.
        Here, we fixed $f_\mathrm{PBH} = 3 \times 10^{-9}$, $c'_1 = 1$ and $c'_2=0.005$.
        The smaller bubbles have larger volume fractions.
        The larger $N_\ast$ becomes, the more lepton bubbles remain.
    }
    \label{fig: vol fraction}
\end{figure}

For $N < N_*$, the probability distribution is the same as in the original setup:
\begin{align}
    \tilde{P}_\mathrm{m}(N,\tilde{\phi} ; c', N_i, \tilde{\phi}_i)
    =
    \tilde{P}(N,\tilde{\phi} ; c', N_i, \tilde{\phi}_i)
    \ ,
\end{align}
where $\tilde{P}_\mathrm{m}$ is the probability distribution in the modified setup.
On the other hand, $\tilde{P}_\mathrm{m}$ for $N \geq N_*$ is obtained by convoluting the $\tilde{P}$ in the first and second stages as
\begin{align}
    \tilde{P}_\mathrm{m}(N, \tilde{\phi}; c'_1, c'_2, N_\ast, N_i, \tilde{\phi}_i)
    =&
    \int \mathrm{d} \tilde{\phi}_\ast \,
    \tilde{P}(N, \tilde{\phi}; c'_2, N_\ast, \tilde{\phi}_\ast)
    \tilde{P}(N_\ast, \tilde{\phi}_\ast; c'_1, N_i,    \tilde{\phi}_i)
    \nonumber\\
    =&
    \int \mathrm{d}\tilde{\phi}_\ast
    \frac{1}{2\pi\sigma^2(N-N_\ast)}
    \exp{\left[
        -\frac{1}{2\sigma^2(N-N_\ast)}
        \Big\lvert
            \tilde{\phi}
            -
            e^{-\frac{c'_2}{2}(N-N_\ast)}\tilde{\phi}_\ast
        \Big\rvert^2
    \right]}
    \nonumber\\
    &\times 
    \frac{1}{2\pi\sigma^2(N_\ast-N_i)}
    \exp{\left[
        -\frac{1}{2\sigma^2(N_\ast-N_i)}
        \Big\lvert
            \tilde{\phi}_\ast
            -
            e^{-\frac{c'_1}{2}(N_\ast-N_i)}\tilde{\phi}_i
        \Big\rvert^2\right]}
    \nonumber\\
    =&
    \frac{1}{2\pi\tilde{\sigma}^2(N)}
    \exp \left[
        -\frac{{\Big\lvert \tilde{\phi}-\tilde{\phi}_i e^{-\frac{c'_1}{2}(N_\ast-N_i) - \frac{c'_2}{2}(N-N_\ast)}\Big\rvert}^2}
        {2\tilde{\sigma}^2(N)}
    \right]
    ,
    \label{eq: itit2}
\end{align}
where
\begin{align}
    \tilde{\sigma}^2(N)
    &\equiv 
    \sigma^2(N-N_\ast,c'_2)
    +e^{-c'_2(N-N_\ast)}\sigma^2(N_\ast-N_i,c'_1)
    \ .
\end{align}
In a similar way to $\bar{P}$, we obtain the probability distribution of $\tilde{\varphi}$ for $N \geq N_*$ as
\begin{align}
\begin{aligned}
    \bar{P}_\mathrm{m}(N,\tilde{\varphi};N_\ast,N_i,\tilde{\varphi}_i)
    =&
    \frac{\tilde{\varphi}}{\tilde{\sigma}^2(N)}
    I_0\left(e^{-\frac{c'_1}{2}(N_\ast-N_i)-\frac{c'_2}{2}(N-N_\ast)}\frac{\tilde{\varphi}\tilde{\varphi}_i}{\tilde{\sigma}^2(N)}\right)\\
    &\times 
    \exp{\left[-\frac{\tilde{\varphi}^2+e^{-c'_1(N_\ast-N_i)-c'_2(N-N_\ast)}\tilde{\varphi}_i^2}{2\tilde{\sigma}^2(N)}\right]}.
    \label{eq: general sol}
\end{aligned}
\end{align}
The condition for lepton bubble formation is that the field value at the end of inflation exceeds the threshold $\tilde{\varphi}_{c}$.
Considering the classical dynamics of the AD field after the horizon exit, the threshold value given at the e-folding number $N$ is obtained as
\begin{align}
    \tilde{\varphi}_{c,\mathrm{eff}}(N)
    =
    \begin{cases}
        \tilde{\varphi}_c e^{\lambda_2(N_\mathrm{end}-N)} 
        & N \geq N_\ast
        \\
        \tilde{\varphi}_c e^{\lambda_2(N_\mathrm{end} - N_\ast) + \lambda_1(N_\ast - N)}            
        & N < N_\ast
    \end{cases}
    \ ,
\end{align}
where
\begin{align}
    \lambda_{1,2}=\frac{3}{2}\left(1 - \sqrt{1-
    \frac{2}{3} c'_{1,2}
    }\right)
    \ .
\end{align}
Note that $\lambda_{1,2} \simeq c'_{1,2}/2$ for $c'_{1,2} \ll 1$.
Using the obtained $\bar{P}_\mathrm{m}$, we can evaluate the PBH abundance and correlation function.
By applying Eq.~\eqref{eq: vol frac} to the modified scenario, we obtain the volume fraction of the high lepton bubbles shown in Fig.~\ref{fig: vol fraction}.

\section{Observational constraints on PBH}
\label{sec: PBH obs}

As mentioned in the introduction, PBH formation from the inhomogeneous AD mechanism can avoid the constraint from the $\mu$-distortion of the CMB spectrum.
However, as for seed PBHs explaining SMBHs, heating by accretion of PBHs changes the ionization history of the universe and affects the CMB anisotropies, from which a stringent constraint is imposed on the PBH abundance as $f_\mathrm{PBH}\lesssim 3\times 10^{-9}$ for $M \sim 10^4 M_\odot$~\cite{Ricotti:2007au,Ali-Haimoud:2016mbv,Poulin:2017bwe,Serpico:2020ehh}.%
\footnote{%
This constraint can be largely relaxed depending on the choice of the accretion model and assumption on the emission efficiency~\cite{Facchinetti:2022kbg}.}
This constraint is accidentally comparable to the estimated SMBH abundance, $f_\mathrm{SMBH} \sim 2.9 \times 10^{-9}$ for $M > 10^6 M_\odot$~\cite{Willott:2010yu} (see also Ref.~\cite{Serpico:2020ehh}).
In this paper, taking into account the ambiguity of the growth of the PBH masses by accretion and mergers, we consider two cases:
$f_\mathrm{PBH}\sim 3\times 10^{-9}$ and $3\times 10^{-10}$.

\begin{figure}[t]
    \centering
    \includegraphics[width=.75\textwidth ]{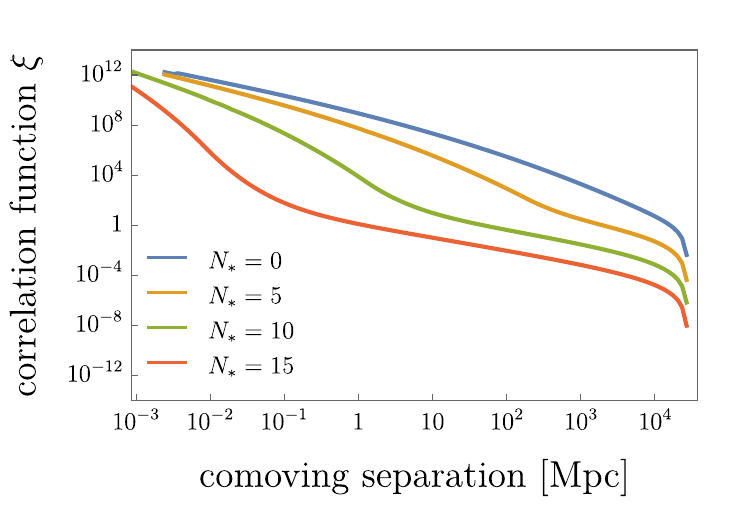}
    \caption{%
        PBH correlation function for different values of $N_*$.
        Here, we take $c'_2=0.005$ and fix $f_\mathrm{PBH}$ by choosing $\tilde{\varphi}_c$.
        The left end has apparent divergence coming from the self-correlation at the scale of the PBH-forming bubble.
        The right end goes to zero at the edge of the observable universe.
        In the actual numerical calculation, it is difficult to evaluate the correlation near those endpoints.
        Larger $N_\ast$ provides the suppression of the correlation in a wider range of scales. 
    }
    \label{fig: correlation}
\end{figure}

In addition, there are constraints on the PBH clustering.
In Fig.~\ref{fig: correlation}, we show the PBH correlation functions in this model.
We fix $f_\mathrm{PBH}=3\times10^{-9}$ for different values of $N_*$ by choosing $\tilde{\varphi}_c=\mathcal{O}(10)$.
Note that $\tilde{\varphi} \gtrsim 2\pi$ is required for the origin, $\varphi=0$, to be stabilized just after inflation.
Here we take $\tilde{c}\simeq d_1\simeq d_2\simeq1$.
The original model corresponds to the case with $N_* = 0$, and the correlation function is more suppressed for larger $N_*$.
In the following, we discuss the constraints from isocurvature perturbations and the angular correlation of quasars.

\subsection{Isocurvature perturbation from PBH clustering}
\label{sec: iso}

\begin{figure}[t]
    \centering
    \includegraphics[width=.75\textwidth ]{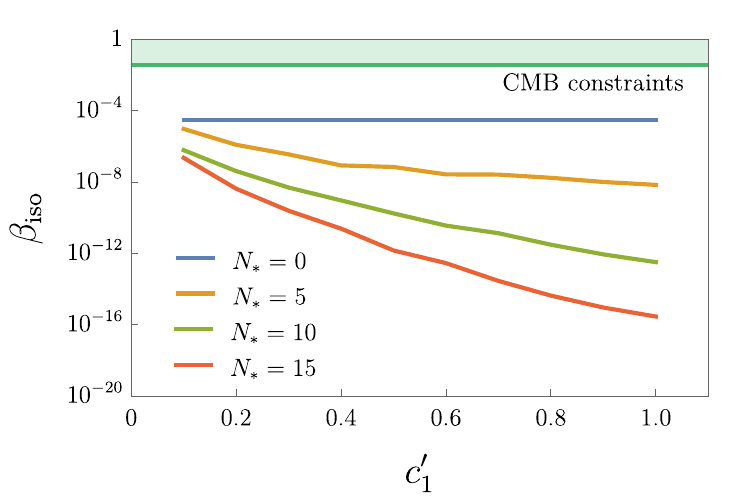}
    \caption{%
        Isocurvature perturbations for the different values of $N_*$.
        Here we take $c'_2=0.005$.
        The green-shaded region shows the excluded region by the CMB observation.
        Even the original model ($N_\ast=0$) is not excluded.
        The modified model has suppressed isocurvature perturbations, and the suppression becomes stronger for larger $N_\ast$.
        Since the fluctuations of the AD field on large scales are suppressed by the Hubble mass in the early stage of inflation, the suppression of $\beta_\mathrm{iso}$ is more significant for larger $c'_1$.
        The results include numerical errors coming from the fitting procedure.
    }
    \label{fig: isocurvature}
\end{figure}

First, we discuss the constraint from isocurvature perturbations on the CMB scales following Refs.~\cite{Kawasaki:2021zir,Kasai:2023ofh}.
We define the power spectrum of the PBH density perturbation by
\begin{align}
    P_\mathrm{PBH}
    =
    \int \mathrm{d}^3\bm{x} e^{- i \bm{k}\cdot \bm{x}}
    \langle \delta_\mathrm{PBH}(0) \delta_\mathrm{PBH}(\bm{x})\rangle 
    =
    \frac{1}{\bar{n}_\mathrm{PBH}}
    +
    P_\xi(k)
    \ ,
\end{align}
where the first term comes from the Poisson fluctuation, and $P_\xi$ is given by 
\begin{align}
    P_\xi(k)
    \equiv 
    \int \mathrm{d}^3\bm{x} \, 
    \xi(x) e^{- i \bm{k}\cdot \bm{x}} 
    =
    4 \pi \int \mathrm{d}r \, 
    r^2 \xi(r) \frac{\sin kr}{kr}
    \ ,
    \label{eq: P_xi integration}
\end{align}
where $r \equiv |\bm{x}|$.
This is tested in terms of the isocurvature perturbations of the CDM density.
The Planck observation puts the constraint~\cite{Planck:2018jri}
\begin{align}
    \beta_\mathrm{iso}
    \equiv\frac{f_\mathrm{PBH}^2\mathcal{P}_\xi(k_\mathrm{CMB})}{f_\mathrm{PBH}^2\mathcal{P}_\xi(k_\mathrm{CMB})+\mathcal{P}_{\mathcal{R}}(k_\mathrm{CMB})}
    <
    0.036
    \ ,
\end{align}
where $\mathcal{P}_\xi \equiv k^3 P_\xi/(2\pi^2)$, and $\mathcal{P}_{\mathcal{R}}(k_\mathrm{CMB}) \simeq 2\times 10^{-9}$ is the dimensionless power spectrum of the curvature perturbations at $k_\mathrm{CMB} = 0.002\,\mathrm{Mpc}^{-1}$~\cite{Planck:2018vyg}.
We show $\beta_\mathrm{iso}$ for different values of $N_*$ in Fig.~\ref{fig: isocurvature}.
The isocurvature perturbation is smaller than the observational constraint even in the original model, and it is suppressed in the modified model.

To evaluate $\beta_\mathrm{iso}$, we simplify the correlation function using a fitting function.
First, we divide the correlation function into the large-scale and small-scale regions at the comoving distance corresponding to $N=N_\ast$.
Then, we fit $\xi(r)$ by power-law functions of $r$ in each region.
Since we expect that $P_\xi(k_\mathrm{CMB})$ is contributed mainly by $\xi(r)$ on large scales, we evaluate only the large-scale contribution to avoid unphysical divergence at $r \sim 0$.
In fact, a similar study for the PBH formation from axion bubbles~\cite{Kasai:2023qic} shows that $P_\xi$ converges by using an analytic expression of $\xi(r)$.
Based on the physical consideration and the result of~\cite{Kasai:2023qic}, 
we expect that the small-scale region does not contribute to the isocurvature perturbation. 

\subsection{Angular correlation functions of SMBH}
\label{sec: ang}

Next, we discuss the angular correlation of the produced PBHs.
We first summarize the definition of the angular correlation function following Ref.~\cite{Shinohara:2021psq}.
We define a two-dimensional PBH number density as the integral of the comoving number density $n_\mathrm{PBH}(R,\theta,\phi)$ over a line of sight,
\begin{align}
    \mathcal{N}_\mathrm{PBH}(\theta,\phi)
    \equiv
    \int_0^\infty \mathrm{d}R \,
    R^2W(R)n_\mathrm{PBH}(R,\theta,\phi)
    \ ,
    \label{eq: 2D number density}
\end{align}
where $\theta$ and $\phi$ are the polar and azimuthal angles of the line of sight, and $W(R)$ is a window function accounting for the range of $R$ relevant to the observations.
The density contrast is defined for $\mathcal{N}_\mathrm{PBH}$ as
\begin{align}
    \Delta_\mathrm{PBH}(\theta,\phi)
    \equiv
    \frac{\mathcal{N}_\mathrm{PBH}(\theta,\phi)-\bar{\mathcal{N}}_\mathrm{PBH}}{\bar{\mathcal{N}}_\mathrm{PBH}}
    \ ,
    \label{eq: 2D number density contrast}
\end{align}
where $\bar{\mathcal{N}}_\mathrm{PBH}$ is the mean value of $\mathcal{N}_\mathrm{PBH}$ over $\theta$ and $\phi$.
For the angular separation $\theta$ between $(\theta_1,\phi_1)$ and $(\theta_2,\phi_2)$, the PBH angular correlation function is defined as~\cite{Shinohara:2021psq}
\begin{align}
    w_\mathrm{PBH}(\theta)
    \equiv &\braket{\Delta_\mathrm{PBH}(\theta_1,\phi_1)\Delta_\mathrm{PBH}(\theta_2,\phi_2)}\\
    = &
    \int_{R_\mathrm{low}}^{R_\mathrm{high}} \mathrm{d}R_1 
    \int_{R_\mathrm{low}}^{R_\mathrm{high}} \mathrm{d}R_2 
    \frac{3R_1^2}{R_\mathrm{high}^3-R_\mathrm{low}^3}
    \frac{3R_2^2}{R_\mathrm{high}^3-R_\mathrm{low}^3} \nonumber \\
    & \times \xi\left( \sqrt{R_1^2+R_2^2-2R_1R_2\cos{\theta}} \right)
    \ ,
    \label{eq: angular correlation}
\end{align}
where $R_\mathrm{high/low}$ denotes the distance to the maximum/minimum redshift of the observations, $z_\mathrm{high/low}$.
We regard the present universe as a flat $\Lambda$CDM one and choose $z_\mathrm{high} = 6.49$ and $z_\mathrm{low} = 5.88$ to match the observations analyzed in Ref.~\cite{Shinohara:2023wjd}.

\begin{figure}[t]
    \centering
    \includegraphics[width=.75\textwidth ]{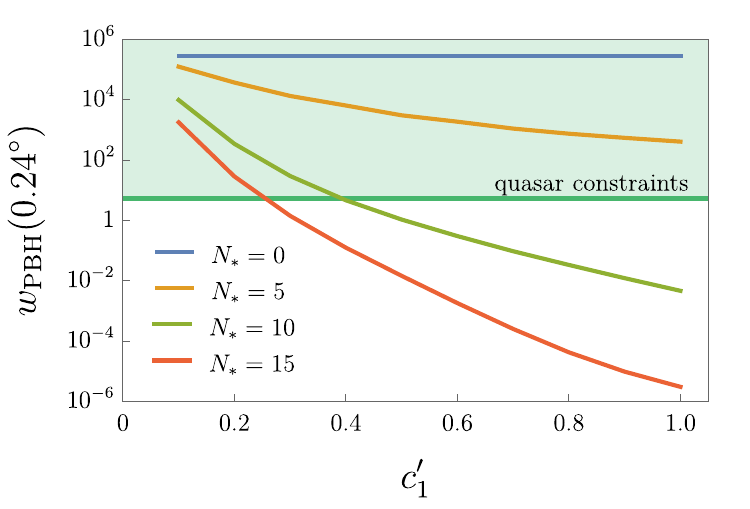}
    \caption{%
        Angular correlation for the different values of $N_\ast$.
        Here we take $c'_2=0.005$.
        The green-shaded region shows the excluded region by the quasar observation.
        The suppression of the angular correlation becomes stronger for larger $N_\ast$ like the isocurvature perturbations. 
        We observe that $N_\ast\geq10$ with $c'_1\geq0.5$ can avoid the observational constraint.
    }
    \label{fig: angular correlation}
\end{figure}

This angular correlation function increases at small angular separations, and then the observational constraint becomes stronger for small $\theta$.
Since $\theta=0.24^\circ$ is the smallest separation in the observational limit, we require that the predicted angular correlation at $\theta=0.24^\circ$ should be smaller than the observational limit.
We show $w_\mathrm{PBH}(0.24^\circ)$ and the observational constraint~\cite{Shinohara:2023wjd} in Fig.~\ref{fig: angular correlation}. 
We find that $0.5\leq c'_1$ and $10 \leq N_\ast$ can avoid the observational constraint from the angular correlation of quasars.

\section{L-ball scenario}
\label{sec: L-ball}

In this section, we explain the properties of L-balls and discuss the observational constraints on the scenario following Ref.~\cite{Kasai:2022vhq}.
In the following, we consider two types of L-balls: new type and delayed type.
First, we review the common properties and then discuss each type.

Although L-balls are classically stable field configurations guaranteed by $U(1)$ symmetry, they finally decay into the standard model lepton charge via quantum effect and thermal effect. 
This can affect both BBN and the effective neutrino number density $N_{\rm{eff}}$, which is constrained by CMB. 
In this section, we investigate the parameter regions consistent with these
constraints.

\subsection{Physics with L-balls}
In gauge-mediated SUSY breaking models, the scalar potential is given by \cite{deGouvea:1997afu}
\begin{align}
    V_\mathrm{gauge}(\phi)
    =
    \begin{cases}
        m_\phi^2{\lvert\phi\rvert}^2 & ({\lvert\phi\rvert}\ll M_S)\\
        M_F^4\left(\ln{\frac{{\lvert\phi\rvert}^2}{M_S^2}}\right)^2 & ({\lvert\phi\rvert}\gg M_S)
    \end{cases}
    \ .
    \label{eq: gauge-med breaking}
\end{align}
Here, $M_F$ and $M_S$ are the mass scales of the SUSY breaking and messenger sectors, respectively.
The gravity mediation effect, although mostly subdominant,  still exists in the gauge-mediated SUSY breaking models and provides the scalar potential~\cite{deGouvea:1997afu}
\begin{align}
    V_\mathrm{grav}(\phi)
    =
    m_{3/2}^2\left[1+K\ln{\frac{{\lvert\phi\rvert}^2}{M_\mathrm{Pl}^2}}\right]{\lvert\phi\rvert}^2
    \ ,
    \label{eq: grav-med breaking}
\end{align}
where $K$ is a dimensionless constant.
Note that the potential~\eqref{eq: grav-med breaking} dominates over the gauge-mediation potential~\eqref{eq: gauge-med breaking} for $\varphi \equiv |\phi| \gtrsim \varphi_\mathrm{eq} \equiv M_F^2/m_{3/2}$.
We assume $K<0$ for the new-type L-ball scenario (Sec.~\ref{sec: new}), and $K>0$ for the delayed-type L-ball scenario (Sec.~\ref{sec: delayed}).

Putting the discussions of the L-ball formation aside, we discuss the time evolution of L-balls here.
Now, we focus on the vacuum B, which forms lepton bubbles.
After the AD field settles to vacuum B, it starts to oscillate when the effective mass becomes comparable to the Hubble parameter.
In this paper, we assume that the oscillation starts when the gravity-mediation effects \eqref{eq: grav-med breaking} dominate the potential. 
Thus, the Hubble parameter and $\varphi_\mathrm{osc}$ at the start of oscillation is given by
\begin{align}
    H_\mathrm{osc}
    &\simeq  m_{3/2}
    \ ,\\
    \label{eq: oscillation Hubble}
    \varphi_\mathrm{osc}
    &\simeq
    \left(
    \frac{\sqrt{\tilde{c}}m_{3/2}M_\mathrm{Pl}^{n-3}}{\lvert\kappa\rvert}
    \right)^\frac{1}{n-2}
    \ .
\end{align}
At this time we neglect the thermal potential~\eqref{eq:thermal_pot} since it is almost subdominant for a parameter region that we are interested in.

After the AD-field oscillation starts, L-balls are formed as discussed later.
L-balls continuously lose and emit their charge due to the thermal effect \cite{Laine:1998rg}.
This process is called evaporation and its rate is written as
\begin{align}
    \Gamma_\mathrm{evap}
    \equiv
    \frac{\mathrm{d}Q}{\mathrm{d}t}\bigg|_\mathrm{evap}
    \sim
    -4\pi\chi\left(\mu_Q-\mu_\mathrm{plasma}\right)T^2R_Q^2
    \simeq -4\pi\chi\omega_QT^2 R_Q^2
    \ ,
    \label{eq: evap rate}
\end{align}
where $Q$ and $R_Q$ are the charge and radius of the L-ball, respectively, and $\mu_{Q,\,(\mathrm{plasma})}$ is the chemical potential for lepton number inside (outside) the L-ball.
Here we use $\mu_Q =\omega_Q$ and $\mu_\mathrm{plasma}\ll \mu_Q$, and adopt the suppression at low temperatures given by~\cite{Laine:1998rg}
\begin{align}
    \chi(T)
    =
    \begin{cases}
        1 & (T>m_\phi)\\
        \left(\frac{T}{m_\phi}\right)^2 & (T<m_\phi)
    \end{cases}
    \ .
    \label{eq: rate coefficient}
\end{align}
For efficient L-ball evaporation, the evaporated charge should be diluted away from the L-ball surface by diffusion~\cite{Banerjee:2000mb}.
The diffusion rate is given by
\begin{align}
    \Gamma_\mathrm{diff}
    \equiv
    \frac{\mathrm{d}Q}{\mathrm{d}t}\bigg|_\mathrm{diff}
    \sim -4\pi DR_Q\mu_QT^2
    \simeq -4\pi DR_Q\omega_QT^2
    \ ,
    \label{eq: diff rate}
\end{align}
where the diffusion constant $D\equiv A/T$ is estimated with a numerical constant $A=4\,\text{--}\,6$. 
When the diffusion rate $\Gamma_\mathrm{diff}$ is smaller than $\Gamma_\mathrm{evap}$, the evaporation process is controlled by diffusion.
Thus, the rate of the evaporation process is determined by the smaller one of $\Gamma_\mathrm{evap}$ and $\Gamma_\mathrm{diff}$, and is given by $\Gamma_\mathrm{diff}$ at high temperatures and $\Gamma_\mathrm{evap}$ at low temperatures.
For later convenience, we define the temperature $\tilde{T}$ for $\Gamma_\mathrm{diff}(\tilde{T})=\Gamma_\mathrm{evap}(\tilde{T})$.
Note that $\tilde{T}<m_\phi$ in the typical parameter region.

The L-balls that survive evaporation finally decay into neutrinos through quantum effects~\cite{Cohen:1986ct}.
The L-ball decay rate is given by~\cite{Kawasaki:2012gk,Kasuya:2012mh}
\begin{align}
    \label{eq: Q-ball_decay}
    \left. \frac{\mathrm{d} Q}{\mathrm{d} t}\right|_\mathrm{decay}
    \simeq 
    - N_\ell \frac{\omega_Q^3}{12\pi^2} 4 \pi R_Q^2
    \ ,
\end{align}
where $N_\ell = 3$ is the number of decay channels.

For later convenience, we clarify the transformation from the cosmic time $t$ to the temperature $T$ here:
\begin{align}
    \frac{\mathrm{d}t}{\mathrm{d}T}
    = 
    \begin{cases}
    -\frac{288}{\pi^2g_\ast}M_\mathrm{Pl}^2H(T_\mathrm{RH})T^{-5}&\left(T\gtrsim T_\mathrm{RH}\right)\\
    -\frac{3}{\pi}\sqrt{\frac{10}{g_\ast}}M_\mathrm{Pl}T^{-3}&\left(T\lesssim T_\mathrm{RH}\right)
    \end{cases}.
\end{align}
\subsection{New-type L-ball}
\label{sec: new}
When $V_\mathrm{grav}(\phi)$ dominates ($\varphi\gtrsim \varphi_\mathrm{eq}$) and $K<0$, ``new-type'' L-balls are formed.
The assumption $\varphi_\mathrm{osc} \gtrsim \varphi_\mathrm{eq}$ is satisfied in most cases of our scenario that explains the PBHs with $10^4 M_\odot$.
The properties of a new-type L-ball are given as follows:
\begin{align}
    Q_N&\simeq\beta_N\left(\frac{\varphi_\mathrm{osc}}{m_{3/2}}\right)^2,\\
    M_Q&\simeq m_{3/2}Q_N,\\
    R_Q&\simeq \lvert K\rvert^{-\frac{1}{2}}m_{3/2}^{-1},\\
    \omega_Q&\simeq m_{3/2},
\end{align}
where $Q_N$ is the charge of the new-type L-ball, and $M_Q$ and $R_Q$ is the mass and radius of the L-ball.
Here the numerical constant $\beta_N$ is estimated by $0.02$ \cite{Hiramatsu:2010dx}.

By substituting new-type L-ball properties to the evaporation and diffusion rate, we can evaluate the lepton charge that evaporates from the L-balls.
A part of the lepton charge is changed to the baryon charge through the sphaleron process.
We are concerned with the baryon charge produced in that way because it can affect the big bang nucleosynthesis (BBN).
Taking into account that the sphaleron process is only effective above the temperature $T_\mathrm{ew}$ at the electroweak phase transition, we finally obtain the total baryon charge evaporating and decaying from an L-ball~\cite{Kasai:2022vhq},
\begin{align}
\begin{split}
    \Delta Q_b^\mathrm{evap}
    \simeq&
    \frac{48M_\mathrm{Pl}}{23}{\lvert K\rvert}^{-1}\Bigg[\frac{\sqrt{10}}{m_\phi^2m_{3/2}}
    \frac{\tilde{T}^2}{\sqrt{g_\ast(\tilde{T})}}
    +2\sqrt{10}A{\lvert K\rvert}^\frac{1}{2}\left(\frac{1}{\sqrt{g_\ast(\tilde{T})}\tilde{T}}-\frac{1}{\sqrt{g_\ast(T_\mathrm{RH})}T_\mathrm{RH}}\right)\\
    &+\frac{64AM_\mathrm{Pl}H(T_\mathrm{RH})}{\pi}{\lvert K\rvert}^\frac{1}{2}
    \frac{1}{g_\ast(T_\mathrm{RH})T_\mathrm{RH}^3}
    \Bigg]\Theta(T_\mathrm{RH}-\tilde{T})\\
    &+\frac{48M_\mathrm{Pl}}{23}{\lvert K\rvert}^{-1}\Bigg[\frac{\sqrt{10}}{m_\phi^2m_{3/2}}
    \frac{T_\mathrm{RH}^2}{\sqrt{g_\ast(T_\mathrm{RH})}}
    +\frac{192M_\mathrm{Pl}H(T_\mathrm{RH})}{\pi m_\phi^2m_{3/2}}\left(\frac{\log{\tilde{T}}}{g_\ast(\tilde{T})} - \frac{\log{T_\mathrm{RH}}}{g_\ast(T_\mathrm{RH})}\right)\\
    &+\frac{64AM_\mathrm{Pl}H(T_\mathrm{RH})}{\pi}{\lvert K\rvert}^\frac{1}{2}
    \frac{1}{g_\ast(\tilde{T})\tilde{T}^3}
    \Bigg]\Theta(\tilde{T}-T_\mathrm{RH})
    \ ,
    \label{eq: new evaporation}
\end{split}
\end{align}
and
\begin{align}
\begin{split}
    \Delta Q_b^\mathrm{decay}
    \simeq
    \frac{576M_\mathrm{Pl}^2m_{3/2}H(T_\mathrm{RH})}{23\pi^3g_\ast(T_\mathrm{RH})T_\mathrm{RH}^4}{\lvert K\rvert}^{-1}
    +\frac{12\sqrt{10}M_\mathrm{Pl}m_{3/2}}{23\pi^2\sqrt{g_\ast(T_\mathrm{ew})}T_\mathrm{ew}^2}{\lvert K\rvert}^{-1}
    \ .
    \label{eq: new decay charge}
\end{split}
\end{align}
Here we have used $T_\mathrm{ew}\ll \tilde{T}$ and $T_\mathrm{RH}\ll T_\mathrm{osc}$ in the parameter region we are interested in.
The temperature when the L-balls decay is estimated by using Eq.~\eqref{eq: new decay charge} as
\begin{align}
    T_\mathrm{decay}
    \simeq
    \left(\frac{3}{2\pi^2}\sqrt{\frac{10}{g_\ast}}{\lvert K\rvert}^{-1}m_{3/2}M_\mathrm{Pl}Q_N^{-1}\right)^\frac{1}{2}
    \ .
    \label{eq: new decay}
\end{align}
\subsection{Delayed-type L-ball}
\label{sec: delayed}
In the case of $K>0$, the ``delayed-type'' L-balls are formed.
The property of a single L-ball is given as follows.
\begin{align}
    Q_G&\simeq\beta_G\frac{\varphi_\mathrm{eq}^4}{M_F^4}=\beta_G\left(\frac{M_F}{m_{3/2}}\right)^4,\\
    M_Q&\simeq \frac{4\sqrt{2}\pi}{3}\zeta M_FQ_G^\frac{3}{4},\\
    R_Q&\simeq \frac{1}{\sqrt{2}\zeta}M_F^{-1}Q_G^\frac{1}{4},\\
    \omega_Q&\simeq \sqrt{2}\pi\zeta M_FQ_G^{-\frac{1}{4}},
\end{align}
where $Q_G$ is the charge of the delayed-type L-ball.
Here the numerical constant $\beta_G\simeq6\times10^{-4}$ and $\zeta\simeq3.6$. 
Similarly to the new-type L-balls, we can evaluate the total baryon charge from an L-ball
as~\cite{Kasai:2022vhq}
\begin{align}
\begin{split}
    \Delta Q_b^\mathrm{evap}
    \simeq&
    \frac{48M_\mathrm{Pl}}{23}
    \Bigg[\frac{\sqrt{5}\pi Q_G^\frac{1}{4}}{\zeta M_Fm_\phi^2}
    \frac{\tilde{T}^2}{\sqrt{g_\ast(\tilde{T})}}
    +2\sqrt{10}A\pi\left(\frac{1}{\sqrt{g_\ast(\tilde{T})}\tilde{T}}-\frac{1}{\sqrt{g_\ast(T_\mathrm{RH})}T_\mathrm{RH}}\right)\\
    &+64AM_\mathrm{Pl}H(T_\mathrm{RH})
    \frac{1}{g_\ast(T_\mathrm{RH})T_\mathrm{RH}^3}
    \Bigg]\Theta(T_\mathrm{RH}-\tilde{T})\\
    &+\frac{48M_\mathrm{Pl}}{23}\Bigg[\frac{\sqrt{5}\pi Q_G^\frac{1}{4}}{\zeta M_Fm_\phi^2}
    \frac{T_\mathrm{RH}^2}{\sqrt{g_\ast(T_\mathrm{RH})}}
    +\frac{96\sqrt{2}M_\mathrm{Pl}H(T_\mathrm{RH})Q_\mathrm{G}^\frac{1}{4}}{\zeta M_Fm_\phi^2}\left(\frac{\log{\tilde{T}}}{g_\ast(\tilde{T})}-\frac{\log{T_\mathrm{RH}}}{g_\ast(T_\mathrm{RH})}\right)\\
    &+64AM_\mathrm{Pl}H(T_\mathrm{RH})
    \frac{1}{g_\ast(\tilde{T})\tilde{T}^3}
    \Bigg]\Theta(\tilde{T}-T_\mathrm{RH})
    \ ,
    \label{eq: delayed evaporation}
\end{split}
\end{align}
and
\begin{align}
\begin{split}
    \Delta Q_b^\mathrm{decay}
    \simeq&
    \frac{8}{23}
    \Bigg[Q_G-\bigg\{Q_G^\frac{5}{4}
    -\frac{15\sqrt{5}\pi\zeta M_FM_\mathrm{Pl}}{4\sqrt{g_\ast(T_\mathrm{ew})}T_\mathrm{ew}^2}
    -
    \frac{90\sqrt{2}\zeta M_FM_\mathrm{Pl}^2H(T_\mathrm{RH})}{g_\ast(T_\mathrm{RH})T_\mathrm{RH}^4}
    \bigg\}^\frac{4}{5}
    \Bigg]
    \ .
    \label{eq: delayed decay charge}
\end{split}
\end{align}
Here, we approximate that the L-ball charge is constant during the evaporation process.
Also, here, we have used $T_\mathrm{ew}\ll \tilde{T}$ and $ T_\mathrm{RH}\ll T_\mathrm{osc}$ in the parameter region we are interested in.
The temperature when the L-balls decay is estimated by using Eq.~\eqref{eq: delayed decay charge} as
\begin{align}
    T_\mathrm{decay}
    \simeq
    \left(\frac{15\pi}{4\sqrt{2}}\sqrt{\frac{10}{g_\ast}}\zeta
    M_\mathrm{Pl}M_FQ_G^{-\frac{5}{4}}\right)^\frac{1}{2}
    \ .
    \label{eq: delayed decay}
\end{align}
\subsection{Observational constraints on L-ball scenario}

For our L-ball scenario to successfully produce the PBHs explaining the SMBH seeds, we consider the following constraints. To obtain the constraints, we take $n=6$ in  $V_{\rm{NR}}$ [Eq.~\eqref{eq : V NR}].
For the observational constraints, we set parameters related to the potential of the AD field, $\tilde{c}=\lvert a_M\rvert=1,~A=5,~\lvert K\rvert=0.1$.
Also we set the energy scales as $M_F=10^7\,\mathrm{GeV},~m_\phi=M_g=10^4\,\mathrm{GeV},~T_\mathrm{ew}=160\,\mathrm{GeV}$~\cite{DOnofrio:2015gop}. 

\begin{itemize}
    \item \textbf{Baryon asymmetry}
    
    In the present model, the spatial distribution of lepton bubbles is highly inhomogeneous.
    Since baryon charge is produced from leptons emitted from L-balls as we calculated in Sec.~\ref{sec: new} and \ref{sec: delayed}, the produced inhomogeneous baryon charge should not affect the homogeneous baryon density not to spoil the success of the BBN.
    Thus, we obtain the constraint as
    \begin{align}
        \braket{\Delta\eta_b}=\eta_l
        \frac{\Delta Q_b^\mathrm{evap}+\Delta Q_b^\mathrm{decay}(T_\mathrm{ew})}{Q_{N,G}}\beta_\mathrm{HLB}\lesssim10^{-2}\eta_b^\mathrm{obs}
        \ ,
    \end{align}
    where $\Delta \eta_b$ denotes the inhomogeneous part of the baryon asymmetry, and $\eta_b^\mathrm{obs} \simeq8.7\times10^{-11}$ denotes the observed value of its homogeneous part~\cite{Planck:2018vyg}.
    $\beta_\mathrm{HLB}$ is the volume fraction of the high lepton bubbles. 
    \item \textbf{Effective number of neutrino generations}
    
    The effective number of neutrino species is calculated as
    \begin{align}
        N_\mathrm{eff}
        =
        \frac{8}{7}\left(\frac{11}{4}\right)^\frac{4}{3}\left[\frac{\rho_\mathrm{rad}}{\rho_\gamma}-1\right]
        \ ,
    \end{align}
    where $\rho_\gamma$ is the energy density of the total photons.
    Neutrinos produced from the L-ball decay could give a significant contribution to $N_\mathrm{eff}$.
    Using the Planck 2018 result~\cite{Planck:2018vyg},
    we can obtain the constraint on the contribution from the L-balls as
    \begin{align}
        \Delta N_\mathrm{eff}
        =
        \frac{8}{7}\left(\frac{11}{4}\right)^\frac{4}{3}\frac{Y_Q^\mathrm{in}s(T_\mathrm{decay})}{\rho_\gamma(T_\mathrm{decay})}\beta_\mathrm{HLB}
        <
        0.24 \ .
    \end{align}
    \item \textbf{Gravitino overproduction}
    
    For the high reheating temperature, gravitinos are over-produced and exceed the present dark matter abundance.
    The gravitino density parameter $\Omega_{3/2}$ is calculated as \cite{Bolz:2000fu, Pradler:2006qh}
    \begin{align}
        \Omega_{3/2}h^2
        \simeq
        0.71\left(\frac{m_{3/2}}{0.5\,\mathrm{GeV}}\right)^{-1}
        \left(\frac{M_g}{10^4\,\mathrm{GeV}}\right)^2
        \left(\frac{T_\mathrm{RH}}{10^5\,\mathrm{GeV}}\right)
        \ ,
    \end{align}
    where $M_g$ is the gluino mass.
    We obtain the constraint requiring that $\Omega_{3/2}h^2$ should not exceed the observed dark matter density $\Omega_\mathrm{DM}h^2\simeq0.12$~\cite{Planck:2018vyg}.
    \item \textbf{PBH formation by L-ball}
    
    Since the present L-ball scenario assumes the PBH formation occurs due to the density perturbation of L-balls, PBH formation must be ahead of the L-ball decay
    \begin{align}
        T_\mathrm{decay}<T_c
        \ .
    \end{align}
    \item \textbf{Dominance of the gravity-mediation potential }

    We assume that the potential due to the gravity-mediation dominates over the gauge-mediation potential when the AD field starts to oscillate.
    Thus we require
    \begin{align}
        \varphi\gtrsim\varphi_\mathrm{eq}=\frac{M_F^2}{m_{3/2}}\gg M_S
        \ .
        \label{eq: new-type condition}
    \end{align}
\end{itemize}

\begin{figure}[t]
    \begin{minipage}[t]{0.45\hsize}
        \centering
        \includegraphics[keepaspectratio, scale=0.6]{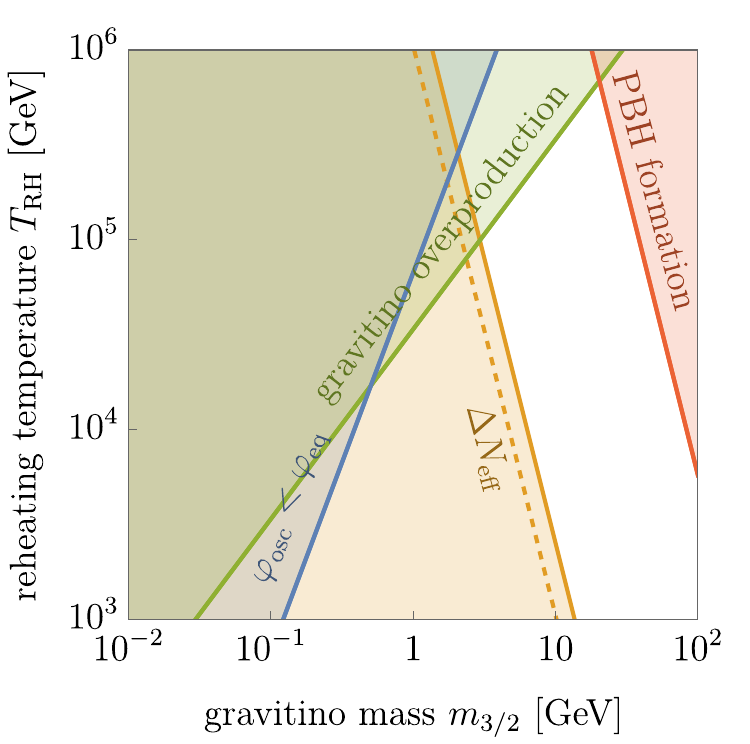}
        \subcaption{new type, $N_\ast=10$}
    \end{minipage}
    \hspace{5mm}
    \begin{minipage}[t]{0.45\hsize}
        \centering
        \includegraphics[keepaspectratio, scale=0.6]{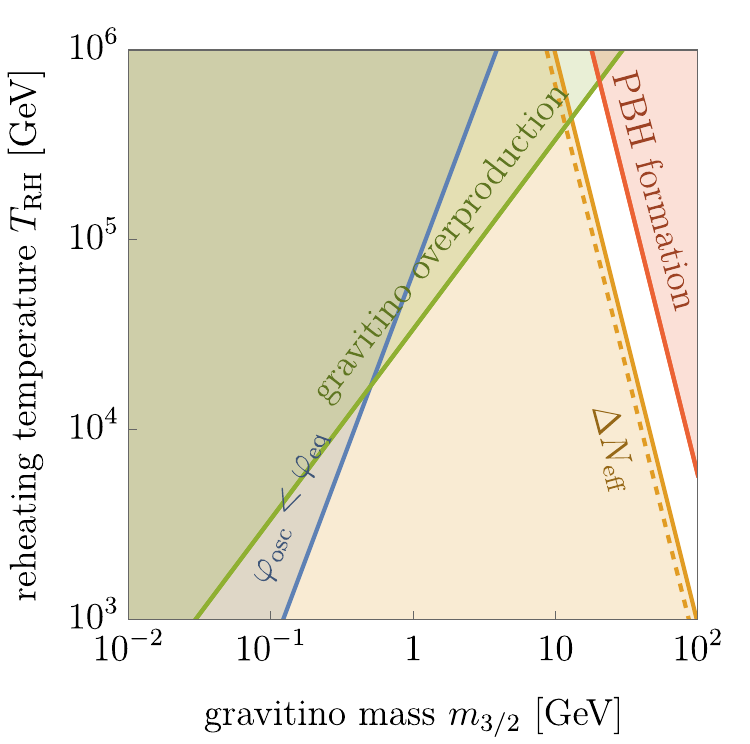}
        \subcaption{new type, $N_\ast=15$}
        \label{Gradation}
    \end{minipage} \\
    \begin{minipage}[t]{0.45\hsize}
        \centering
        \includegraphics[keepaspectratio, scale=0.6]{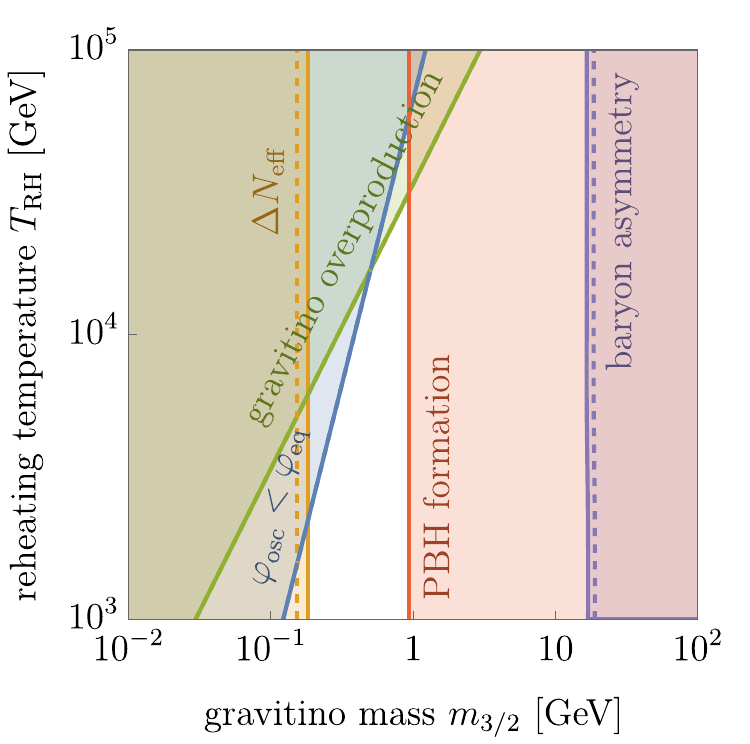}
        \subcaption{delayed type, $N_\ast=10$}
    \end{minipage}
    \hspace{5mm}
    \begin{minipage}[t]{0.45\hsize}
        \centering
        \includegraphics[keepaspectratio, scale=0.6]{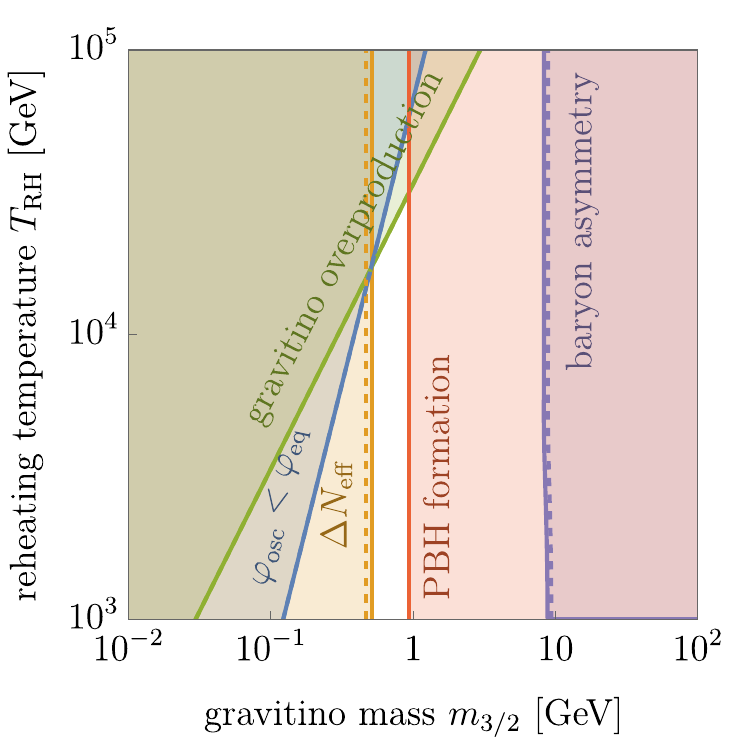}
        \subcaption{delayed type, $N_\ast=15$}
    \end{minipage}
    \caption{%
        Observational constraints on the L-ball scenario with $c'_1=0.5$.
        Here we take $c'_2=0.005$.
        The solid line shows the constraints for $f_\mathrm{PBH}=3\times 10^{-9}$, and the dashed line shows that for $f_\mathrm{PBH}=3\times 10^{-10}$ only if they are different.
    }
    \label{fig: constraints c=0.5}
  \end{figure}

\begin{figure}[t]
    \begin{tabular}{cc}
      \begin{minipage}[t]{0.45\hsize}
        \centering
        \includegraphics[keepaspectratio, scale=0.6]{c=1,N=10,new.pdf}
        \subcaption{new type, $N_\ast=10$}
        \label{composite_c1}
      \end{minipage} 
      \hspace{5mm}
      \begin{minipage}[t]{0.45\hsize}
        \centering
        \includegraphics[keepaspectratio, scale=0.6]{c=1,N=15,new.pdf}
        \subcaption{new type, $N_\ast=15$}
        \label{Gradation_c1}
      \end{minipage} \\
   
      \begin{minipage}[t]{0.45\hsize}
        \centering
        \includegraphics[keepaspectratio, scale=0.6]{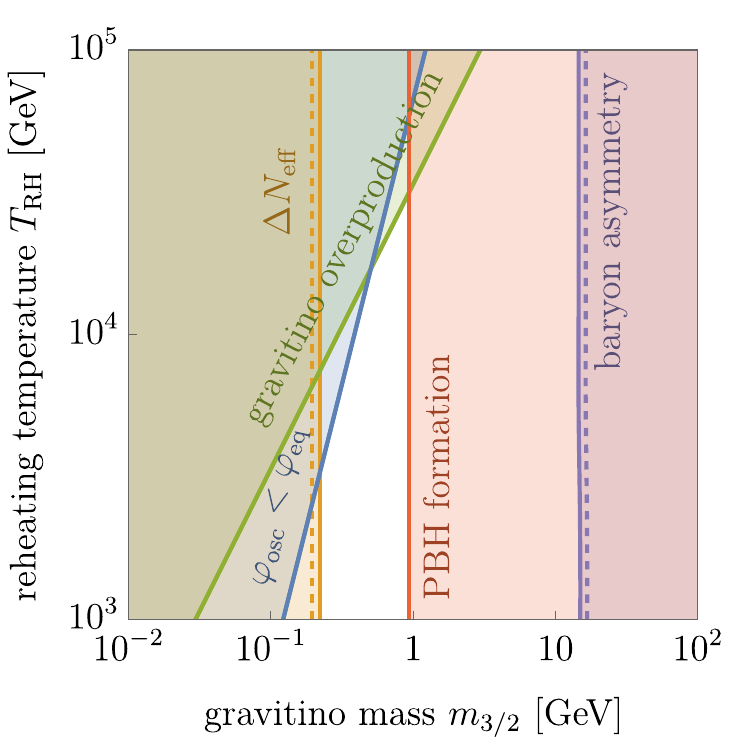}
        \subcaption{delayed type, $N_\ast=10$}
        \label{fill_c1}
      \end{minipage} 
      \hspace{5mm}
      \begin{minipage}[t]{0.45\hsize}
        \centering
        \includegraphics[keepaspectratio, scale=0.6]{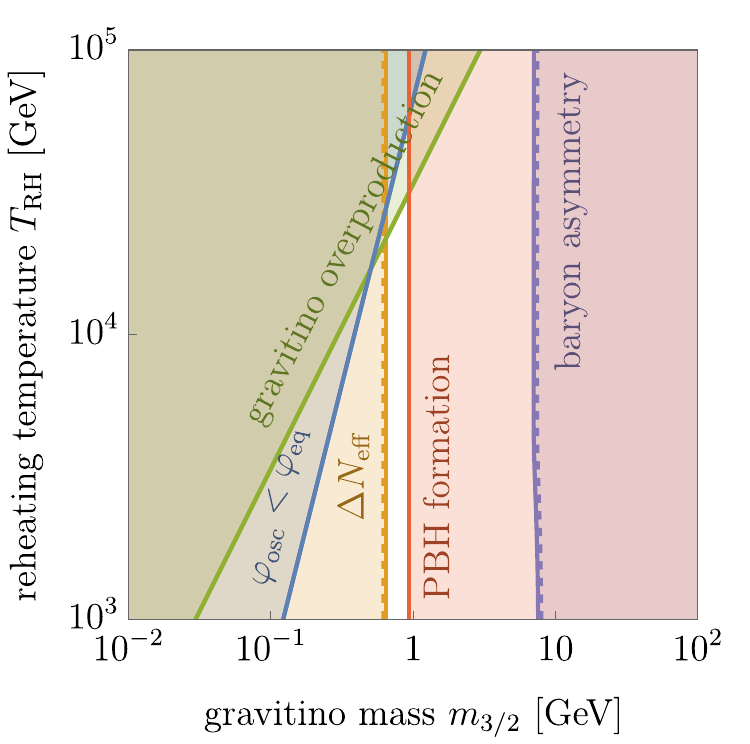}
        \subcaption{delayed type, $N_\ast=15$}
        \label{transform_c1}
      \end{minipage} 
    \end{tabular}
    \caption{%
        Same as Fig.~\ref{fig: constraints c=0.5}, but for $c'_1=1.0$.
    }
    \label{fig: constraints c=1.0}
  \end{figure}

We focus on $c'_1 \geq 0.5$ and $N_\ast \geq 10$ considering the constraints on the PBH correlation discussed in Sec.~\ref{sec: PBH obs}.
We show the parameter region for the L-ball scenario for $c'_1 = 0.5$ and $1.0$ in Figs.~\ref{fig: constraints c=0.5} and \ref{fig: constraints c=1.0}, respectively.
Here, we fix $f_\mathrm{PBH}=3\times10^{-9}$ and show the constraints listed above.
For the constraints depending on $f_\mathrm{PBH}$, we show those for $f_\mathrm{PBH}=3\times10^{-10}$ with the dashed lines.
All these cases are stringently constrained but we have allowed parameter regions.
%

\section{Summary and discussion}
\label{sec: summary}

In this paper, we have studied a modified scenario to form PBHs with masses of $M \gtrsim 10^4 M_\odot$ via the inhomogeneous AD leptogenesis motivated by the observations of SMBHs.
In this type of model, the PBH formation depends on quantum fluctuations of the AD field during inflation, and the formed PBHs inevitably have strong spatial correlations, which are stringently constrained on large scales by observations of CMB and the angular correlation of quasars.
By considering a time-dependent Hubble mass during inflation, which can be realized, for example, in multi-inflaton models, we can suppress the spatial correlation of PBHs on large scales, and the constraints from isocurvature perturbations and angular correlation of quasars are relaxed.

This scenario accompanies the formation of L-balls.
The L-balls source the density contrast for PBH formation and protect the lepton asymmetry against the sphaleron processes.
Consequently, we can successfully explain the seeds of the SMBHs without conflicting with the observational constraints.

While the PBH correlation is suppressed on large scales, it can have some implications on small scales.
For example, the clustered PBHs will result in the clustering of SMBHs on smaller scales than observed today.
It may affect the radiation of gravitational waves from SMBH binaries. 

In addition, small bubbles do not collapse into PBHs and result in rare regions with large lepton asymmetry after the L-ball decay.
The energy of the L-balls is emitted as relativistic neutrinos and does not contribute to the isocurvature perturbations.
On the other hand, if the L-balls decay before the BBN at $T \sim \mathcal{O}(1)$\,MeV, the abundance of light elements can be altered in such regions.
In particular, large lepton asymmetry significantly affects the abundance of $^4$He.
Thus, this scenario may predict enhanced or suppressed abundance of $^4$He for galaxies in the rare regions, which is left for future study.

Finally, we comment on the assumption in our analysis.
For simplicity, we have assumed a constant Hubble parameter during inflation.
However, the Hubble parameter typically decreases during inflation.
In particular, the Hubble parameter can significantly decrease at the transition of the stages in multi-inflaton models.
If we take into account the time evolution of the Hubble parameter, the correlations on large scales are enhanced because the field fluctuations in the earlier period become relatively larger.
This enhancement, however, can be compensated by $c_{1, 2}$.
Thus, we expect our conclusion to be unchanged by the time-dependent Hubble parameter.

\begin{acknowledgments}
This work was supported by JSPS KAKENHI Grant Nos. 20H05851 (M.K.), 21K03567 (M.K.), 23KJ0088 (K.M.), and 24K17039
(K.M.).
This work was supported by JST SPRING, Grant Number JPMJSP2108 (K.K. and S.N.).
\end{acknowledgments}

\bibliographystyle{JHEP}
\bibliography{Ref}

\end{document}